# Highly efficient multi-chromatic Raman microlasers from cavity polygon modes on thin-film lithium niobate platform

YIXUAN YANG,[1,4] CHUNTAO LI,[2,3] RENHONG GAO,[2] YINGNUO QIU,[1,4] LINGLING QIAO,[1] JIELEI NI,[5, †] JINTIAN LIN,[1,4,*] AND YA CHENG[2,3,6,7,8,9,‡]

[1]*State Key Laboratory of Ultra-intense laser Science and Technology, Shanghai Institute of Optics and Fine Mechanics, Chinese Academy of Sciences, Shanghai 201800, China*
[2]*The Extreme Optoelectromechanics Laboratory (XXL), School of Physics and Electronic Science, East China Normal University, Shanghai 200241, China*
[3]*State Key Laboratory of Precision Spectroscopy, East China Normal University, Shanghai 200062, China*
[4]*Center of Materials Science and Optoelectronics Engineering, University of Chinese Academy of Sciences, Beijing 100049, China*
[5]*Nanophotonics Research Center, Institute of Microscale Optoelectronics & State Key Laboratory of Radio Frequency Heterogeneous Integration, Shenzhen University, Shenzhen 518060, China*
[6]*Shanghai Research Center for Quantum Sciences, Shanghai 201315, China*
[7]*Hefei National Laboratory, Hefei 230088, China*
[8]*Collaborative Innovation Center of Extreme Optics, Shanxi University, Taiyuan 030006, China*
[9]*Collaborative Innovation Center of Light Manipulations and Applications, Shandong Normal University, Jinan 250358, China*
[†]*jieleini@yeah.net*
[*]*jintianlin@siom.ac.cn*
[‡]*ya.cheng@siom.ac.cn*

**Abstract:** The integration of stimulated Raman scattering (SRS) and second order nonlinearity ($\chi^{(2)}$) in non-centrosymmetric photonic microresonators presents a highly promising solution for developing on-chip coherent light sources with exceptional bandwidth and flexible tunability. Our study introduces an innovative methodology leveraging cavity polygon modes within an X-cut thin-film lithium niobate microdisk to achieve highly efficient multi-chromatic Raman microlasers. Specifically, high-Q square modes characterized by two parallel sides oriented perpendicularly relative to the optical axis of lithium niobate crystal were excited. These modes offer distinct advantages, including enhancing both mode-field overlap and improved phase matching, achieved through the utilization of the largest second-order susceptibility component ($d_{33}$), which is critical for Raman-quadratic nonlinear interactions. The experimental results highlight significant advancements in multi-wavelength multi-wavelength laser generation, with forward stimulated Raman laser signals exhibiting a high conversion efficiency of up to 65.02% and an impressively narrow integral linewidth of only 5.2 kHz. Simultaneously, our system enables the generation of multi-wavelength Raman-quadratic laser signals across the ~800 nm and ~530 nm spectral bands. These findings are further underscored by an impressive absolute conversion efficiency of 1.33% for the 797.4-nm Raman laser, achieved at a remarkably low pump power of just 1.07 mW. This work not only extends the application scope of cavity polygon modes from single second/third-order nonlinear optical processes to cascaded processes but also establishes a foundation for realizing high-efficiency on-chip multi-chromatic laser sources with versatile functionalities.

## 1. Introduction

Nonlinear optical processes, including second harmonic generation (SHG), optical parametric down-conversion (OPDC), four-wave mixing (FWM), stimulated Raman scattering (SRS), and stimulated Brillouin scattering (SBS), are pivotal in expanding the bandwidth of light sources and enabling the engineering of novel light sources [1-9]. These processes not only play a crucial role in advancing fundamental scientific research but also hold immense potential for applications across various fields such as quantum technology, precision metrology, data processing, and sensing across various fields [1-9]. To enhance nonlinear interactions, non-centrosymmetric optical crystals with broad transparency windows, particularly thin-film lithium niobate (TFLN), have been extensively utilized, offering high second-order nonlinearity and strong light field confinement, and yielding highly efficient nonlinear conversion [10-19]. Recent advancements in TFLN-based nonlinear photonics have sparked interest in integrating two or more nonlinear processes to extend coherent spectral bandwidths [18-26]. This integration aims to enhance on-chip functionalities such as self-referencing supercontinuum generation, bi-chromatic soliton microcombs, synthesized multi-wave mixing, and cascaded third harmonic generation. However, compared to individual nonlinear processes, the incorporation of multiple nonlinear processes presents unique challenges, including maintaining mode field overlap, satisfying phase matching conditions, and minimizing device losses. Despite leveraging high-Q whispering gallery modes (WGMs) in TFLN microresonators to enhance nonlinear interactions, achieving high efficiency in multi-process integration remains a significant challenge.

Recently, polygon modes have emerged as a promising alternative to conventional WGMs in TFLN microdisk resonators [16,27-31]. By introducing weak perturbations through a coupled tapered fiber, light can be coherently recombined to form polygon-shaped orbits within the microdisk, propagating along trajectories that are largely isolated from the rough cavity edges. This unique spatial distribution offers three significant advantages: ultra-high intrinsic Q factors exceeding $10^7$ due to reduced scattering losses, high modal overlap factors (~80%) between interacting waves, and agile post-fabrication dispersion management [16]. Leveraging these advantages, polygon modes have achieved remarkable success in single second/third-order nonlinear processes, particularly SHG with 48% absolute conversion efficiency [30] and soliton microcomb generation with a low threshold of 11 mW [16]. However, whether polygon modes can provide similar enhancement for cascaded nonlinear processes to extend on-chip coherent spectral coverage, which involve both second-order ($\chi^{(2)}$) and third-order ($\chi^{(3)}$) nonlinear processes, remains an open question. Moreover, the unique geometry of polygon modes in X-cut TFLN microresonators enables broadband quasi-phase matching without sophisticated periodic poling, which is particularly advantageous for cascaded nonlinear processes involving multiple wavelengths.

Here, we demonstrate, for the first time, that polygon modes provide significant enhancement for cascaded nonlinear processes, thereby establishing a complete nonlinear optical framework encompassing both $\chi^{(2)}$ and $\chi^{(3)}$ processes in a single microcavity platform. Circular microdisk resonators with ultra-smooth surfaces were fabricated via photolithography assisted chemo-mechanical etching on X-cut TFLN wafers to suppress scattering loss. By introducing a weak perturbation in the microdisk using a coupled tapered fiber, high-Q square modes were excited with two parallel sides perpendicular to the optical axis of lithium niobate crystal. This configuration enables leveraging the highest second-order susceptibility component ($d_{33}$) of lithium niobate to enhance quadratic nonlinear interactions. Meanwhile, the strict confinement of square modes by classical orbits ensures high mode field overlap for efficient nonlinear processes. When a square mode at 1566.6 nm was excited, we achieved multi-wavelength forward Stokes stimulated Raman laser (SRL) with high conversion efficiency. Furthermore, the experimental results demonstrate the potential of polygon modes in enabling cascaded Raman-quadratic nonlinear processes, paving the way for advanced on-chip photonic applications.

## 2. Generation of highly efficient multi-chromatic Raman microlasers on cavity polygon modes

*2.1 Fabrication of the TFLN microdisk resonators*

A commercially available 700 nm-thick X-cut lithium niobate on insulator wafer (NanoLN) was used to fabricate the microdisks for generating multi-chromatic Raman microlasers. The microdisks were produced via femtosecond laser photolithography-assisted chemo-mechanical etching (PLACE) technique with an ultra-smooth surface through six steps [19,32]. First, a chromium (Cr) layer was deposited on the wafer. Subsequently, the Cr layer was subtractively ablated using a focused femtosecond pulsed laser to form microdisk-shaped patterns with a high spatial resolution of 200 nm. Third, the microdisk patterns were transferred from the Cr layer to the TFLN through chemo-mechanical polishing (CMP). Fourth, chemical etching was employed to remove the Cr patterns. Fifth, secondary CMP was performed to improve the surface smoothness of the TFLN microdisks and reduce the microdisk thickness to 0.65 μm. Finally, the silica layer beneath the TFLN microdisks was etched to form small pedestals for supporting the suspended TFLN microdisks. The fabricated microdisks possess a diameter of 55.67 μm, as depicted in the inset of Fig. 1.

## 2.2 Experimental setup

Figure 1 shows the schematic diagram of the experimental setup for realizing high-conversion-efficiency multi-chromatic Raman microlasers on cavity polygon modes. A tunable diode laser operating in the telecom C band was selectively tuned to be transverse-electrically (TE) polarized via an inline fiber polarization controller (PC), and injected into an erbium-doped fiber amplifier (EDFA) and a variable optical attenuator (VOA) for power amplification and regulation. Then the light was input into a tapered fiber with a waist of 2 μm, which was in direct contact with the top surface of the microdisk and inclined to the optical axis of the lithium niobate crystal at an angle of 45°. This coupled configuration simultaneously enables coupling light into and out of the microdisk, and inducing weak perturbation to the microdisk for the excitation of square modes with two parallel sides perpendicular to the optical axis of lithium niobate. An optical microscope imaging system was built above the microdisk, consisting of a microscope objective with a numerical aperture (NA) of 0.28, optical filters, and an infrared/visible charge-coupled device (CCD) camera, for real-time monitoring of the microdisk-fiber coupling position. The output signals from the tapered fiber were sent to an optical spectrum analyzer (OSA) and a fiber spectrometer for spectral analysis. For the characterization of the Q factors of the cavity modes, the output signals were directed to a photodetector (PD) connected with an oscilloscope. Meanwhile, a signal generator was used to linearly scan the wavelength of the tunable diode laser across the resonant wavelength of the cavity modes, and the transmission spectrum of the tapered fiber coupled with the microdisk was recorded by the oscilloscope during the wavelength scanning process.

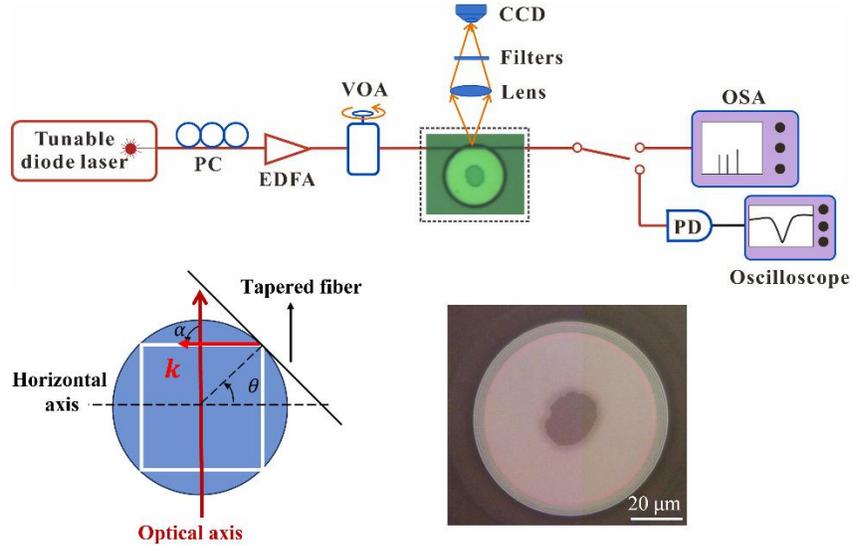

Fig. 1. Schematic diagram of the experimental setup for realizing high-efficiency multi-chromatic Raman microlasers on cavity square modes in the microdisk. Inset (left bottom): schematic of the excitation of the square modes in the X-cut microdisk. Inset (right bottom): Optical microscope image of the fabricated microdisk.

### *2.3 Ultra-efficient SRS processes*

When the pump laser wavelength was set to 1566.6 nm ($\lambda_p$) with a pump power exceeding the SRS threshold, four distinct Raman laser signals were observed at 1624.2 ($\lambda_{R-I-1}$), 1733.5 ($\lambda_{R-I-2}$), 1806.8 ($\lambda_{R-II-1}$), and 1815.6 nm ($\lambda_{R-II-2}$), as shown in Fig. 2(a). The infrared light emission from the microdisk surface was recorded by the optical microscope imaging system, as depicted in the inset of Fig. 2(b). The recorded intensity profile shows a square-shaped pattern, which significantly differed from the WGM characteristic.

The strongest Raman laser signal at 1624.2 nm was generated via the SRS process involving an optical phonon branch [E(TO)$_3$], corresponding to a Raman frequency shift of 238 cm$^{-1}$. The Raman laser signal at 1733.5 nm was also excited by the pump light with a Raman frequency shift of 613 cm$^{-1}$. The other two Raman laser signals at 1806.8 nm and 1815.6 nm, were generated through cascaded SRS processes, utilizing the SRS signals at 1733.5 nm as the pump source. These cascaded signals exhibited frequency shifts of ~239 cm$^{-1}$, and ~276 cm$^{-1}$ [33], respectively. Figure 2(b) illustrates the output power of the SRS signal at 1624.2 nm as a function of on-chip forward pump power. When the pump power exceeded a certain threshold value, the output power of the Raman laser exhibited a linear increase with respect to the pump power. Based on the linear fitting analysis, the SRS signal was determined to have a threshold of 0.50 mW. The conversion efficiency of the Raman laser reached 65.02%, which represents a state-of-the-art performance in the TFLN platform [25,26,34]. This high conversion efficiency is attributed to the exceptional mode field overlap factor between the pump and SRS modes (93.81%), as calculated using the finite element method [28], as shown in the inset of Fig. 2(c). When the pump power was increased to 1.07 mW, the maximum peak power achieved was 215 μW, with a side-mode suppression ratio as high as 44.3 dB, as demonstrated in Fig. 2(c).

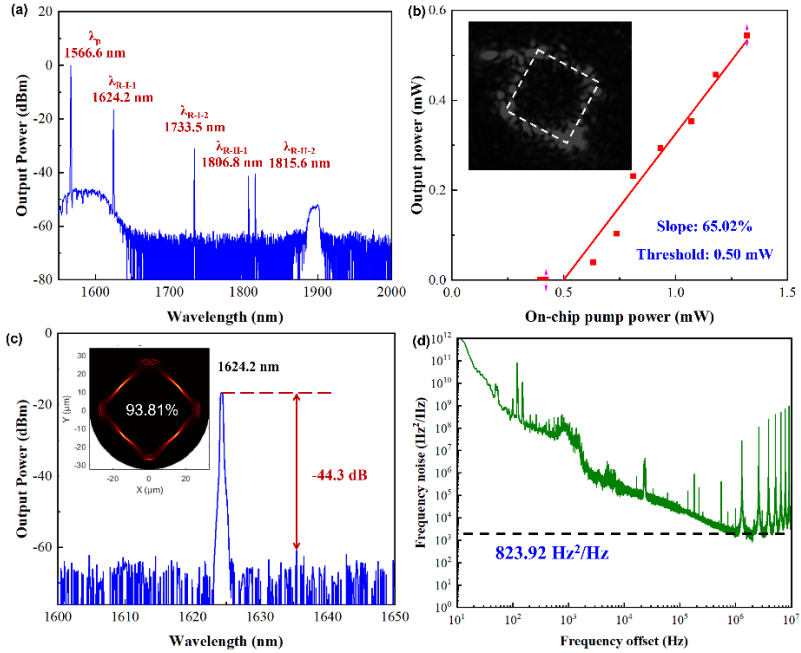

Fig. 2. (a) Spectrum of the generated SRS signals. (b) Output power of the SRS signal at 1624.2 nm wavelength as a function of the on-chip forward pump power. Inset: Intensity profile of the infrared light emission from the microdisk surface. (c) The sidemode suppression ratio of the SRS signal at 1624.2 nm. Inset: calculated mode field overlap factor of the pump and SRS square modes. (d) The frequency noise spectrum of the SRS signal at 1624.2 nm.

The integral linewidth of the generated Raman laser at 1624.2 nm was further characterized using a commercial laser phase noise analyzer based on the correlated self-heterodyne method [35-37], with an integration time of 1 ms. As shown in Fig. 2(d), the frequency noise spectrum reveals a white-frequency-noise floor of 823.92 Hz²/Hz and demonstrates short-term linewidths as narrow as 5.176 kHz.

## 2.4 Highly-efficient Raman-quadratic lasing and harmonics generation processes

Upon further increasing the pump power to 1.07 mW, visible-light emission from the microdisk became clearly observable under naked-eye inspection and captured by the optical microscope imaging system, as shown in Fig. 3(a). The intensity profile of the visible-light emission exhibited a distinct square-patterned characteristic. Meanwhile, an optical spectrum in 350–1000 nm was recorded by the OSA and fiber spectrometer, as depicted in Fig. 3(e). In this spectrum, four short-wavelength signals near 800 nm were identified at specific wavelengths: 783.3 ($\lambda_{SHG}$), 797.4 ($\lambda_{RQ-I-1}$), 822.9 ($\lambda_{RQ-I-2}$), and 840.9 nm ($\lambda_{RQ-I-3}$). Based on the energy conservation, the signal at 783.3 nm was formed through second harmonic generation (SHG) of the pump light, following the relation $1/\lambda_{SHG}=2/\lambda_p$. The strongest near-infrared signal observed at 797.4 nm ($\lambda_{RQ-I-1}$) resulted from a Raman-quadratic process, specifically sum frequency generation (SFG) involving the pump light and the 1624.2-nm SRS signal, as indicated by $1/\lambda_{RQ-I-1}=1/\lambda_p+1/\lambda_{R-I-1}$. Similarly, the other signals at 822.9 nm and 840.9 nm) were also produced through SFG processes between the pump light and the 1repective SRS signals at 733.5 nm and 1815.6 nm). To observe the intensity profiles of these four 800-nm vicinity signals within the microdisk, two filter configurations were employed in the optical microscope imaging system. The first configuration utilized an 800 nm long-pass filter, while the second combined an 800 nm short-pass filter with a 700 nm long-pass filter. The resulting emission patterns are presented in Figs. 3(b) and 3(c), respectively. Notably, all four signals displayed square-shaped intensity profiles, deviating from the conventional whispering gallery mode characteristics. The measured output power of the second harmonic signal at 1.07 mW pump level was 10.12 μW, corresponding to an absolute efficiency of 0.71%. The strongest signal at 797.4 nm exhibited a significantly higher absolute efficiency of 1.33%, which can be attributed to favorable phase matching conditions and high modal overlap factors exceeding 80%.

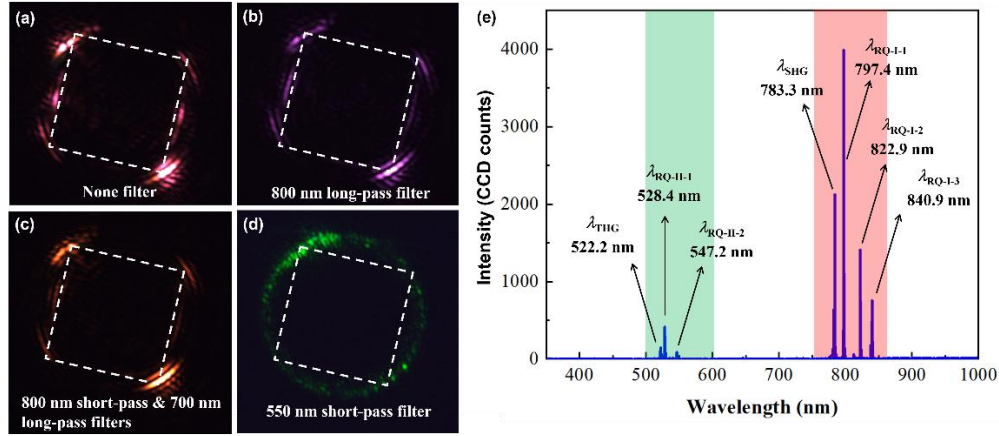

Fig. 3. (a)-(d) Scattered light from the microdisk surface captured by the optical imaging system with some filters. (e) Spectrum of Raman-quadratic laser signals and harmonics signals.

Additionally, three signals with relatively weak powers were observed at shorter wavelengths: 522.2 ($\lambda_{THG}$), 528.4 ($\lambda_{RQ-II-1}$), and 547.2 nm ($\lambda_{RQ-II-2}$). Based on the principle of energy conservation, the generation mechanisms of these signals can be explained as follows: The 522.2-nm signal was formed through a cascaded third harmonic generation (cTHG) process, specifically the SFG process between the pump light and the second harmonic, satisfying the relation $1/\lambda_{THG}=1/\lambda_p+1/\lambda_{SHG}$. The 528.4-nm signal was generated via SFG of the pump light and the 797.4-nm signal ($1/\lambda_{RQ-II-1}=1/\lambda_p+1/\lambda_{RQ-I-1}$), while the 547.2-nm signal resulted from SFG of the pump light and the 840.9-nm signal ($1/\lambda_{RQ-II-2}=1/\lambda_p+1/\lambda_{RQ-I-3}$). As shown in Fig. 3(d), an optical microscope image captures the visible-light emission from the microdisk surface, revealing green light emission characterized by a square-mode characteristic and accompanied by high-order WGM features. Notably, the absolute conversion efficiency of the strongest signal at 528.4 nm in this visible waveband was measured to be 0.38%, demonstrating significant potential for practical applications.

*2.5 Analysis of the underlying physical mechanism behind the highly efficient nonlinear light sources*

Typically, for $n^{th}$-order nonlinear interactions in whispering gallery microresonators, the conversion efficiency is proportional to a parameter $Q^n/V$. Here, Q represents the quality factor of the WGMs, while V denotes the mode volume which scales with the physical dimensions of the microresonators [1-7,38,39]. In our specific microresonator design, the loaded Q factor of the pump mode achieved $2.6\times10^6$, whereas the loaded Q factor near 781 nm was measured to be $1.0\times10^6$, as shown in Fig. 4. Previous studies have revealed that nonlinear interactions on polygon modes of almost the same patterns in the weakly perturbed microdisks benefit from high mode field overlap factors of ~80%, despite the dispersion effect [16,28]. The combination of these high Q factors, the compact size of the microresonator, and the significant mode field overlap ensures substantial cavity built-up energy flux densities. This enables both optical parametric and non-parametric processes with high conversion efficiencies. Consequently, SRS processes can be effectively excited on square modes in this high-Q LNOI microdisk platform, achieving record-high conversion efficiencies within the LNOI framework. These multi-wavelength SRS processes further provide stimulated Raman lasing signals for subsequent Raman-quadratic processes.

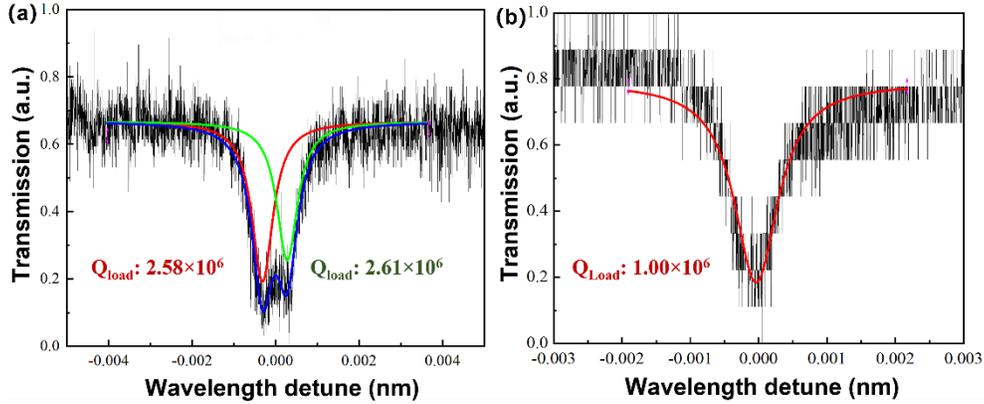

Fig. 4. (a) Loaded Q factor of the pump mode. (b) Loaded Q factor around 781nm.

Furthermore, in optical parametric processes such as harmonic generation and Raman-quadratic processes within lithium niobate microresonators, achieving phase matching is essential. Considering the unique second-order nonlinear susceptibilities of lithium niobate crystals, it is particularly advantageous to leverage the significantly higher $d_{33}$ coefficient, which exceeds other susceptibility components by an order of magnitude. This makes $d_{33}$ utilization critical for high-efficiency processes like SHG and SFG. The conventional method for exploiting $d_{33}$ involves creating periodically poled microdomains within the lithium niobate crystal structure. While this approach can achieve high absolute conversion efficiency, it necessitates complex fabrication techniques, including precise microelectrode lithography and high-voltage poling processes to form fine poled domains with well-defined domain walls. These manufacturing steps not only increase production complexity and costs but also introduce scattering losses and limit the quasi-phase matched bandwidth. To enable the simultaneous generation of multi-chromatic Raman-quadratic lasing within a single microresonator, an advanced phase matching scheme capable of supporting broad bandwidth operation is required.

In this work, transverse-electrically (TE) polarized cavity modes were excited with square-shaped intensity profile of two sides perpendicular to the optical axis of lithium niobate crystal in the dispersion engineered X-cut LNOI microdisk. The effective second-order nonlinear coefficient can be expressed as [30]

$$d_{eff} = -d_{22}cos3\alpha + 3d_{31}cos2\alpha sin\alpha + d_{33}sin3\alpha \approx d_{33}sin3\alpha, \quad (1)$$

where $\alpha$ is the intersection angle between the wave vector $\boldsymbol{k}$ of the TE polarized wave and the optical axis. Assuming that the light field within the square pattern is uniform, under slowly varying amplitude approximation, the growth rate of the field amplitude $E_{SFG}$ of the SFG signal at the propagated path $l$ along the square pattern can be expressed as [30]

$$E_{SFG}(l) = \frac{2id_{eff}\omega_{SFG}^2 E_p E_{p\prime}}{k_{SFG}c^2} \int_0^L e^{i\Delta k(l)\cdot l}dl. \quad (2)$$

Here, $\omega_{SFG}$, $E_p$, $E_{p\prime}$, $k_{SFG}$, $c$ and $L$ are angular frequency of the SFG signal, field amplitude of the pump light, field amplitude of the other input light participating in the SFG, wave vector of the SFG signal, speed of light in vacuum, and the physical length of the side of the square pattern. As an example, we take the cascaded Raman-quadratic processes for the generation of 797.4-nm and 528.4 nm lasing signals, to expand this phase matching scheme from the previously reported narrow-bandwidth phase matched SHG to the broad-bandwidth cascaded $\chi^{(2)}$ process for the first time. For the generation of the 797.4-nm Raman-quadratic lasing signal, the magnitude of this SFG signal significantly increases with azimuthal angle $\theta$ sections of $\pi/2–\pi$ and $3\pi/2–2\pi$ in every propagated cycle, as shown in Fig. 5(a). Consequently, there are 2 stepped growths in every cycle, enabling the buildup of the Raman-quadratic lasing signal in the high-Q microdisk. Additionally, in each growth section, the Raman-quadratic lasing signal oscillates with 99 coherent buildup lengths as light propagates along 1/4 cavity-cycle, as depicted in Fig. 5(b), leading to a net gain in the entire growth sections. And the high-Q factor of the microdisk will enhance the nonlinear interaction length for yielding a highly efficient Raman-quadratic lasing signal.

For the generation of the 528.4-nm cascaded Raman-quadratic lasing signal, it was formed by the cascaded SFG of the pump light and the generated 797.4-nm signal. Akin to the growth dynamics of the SFG, this cascaded SFG signal also undergoes remarkable gain with azimuthal angle $\theta$ sections of $\pi/2–\pi$, and $3\pi/2–2\pi$ in each propagation cycle, as shown in Figs. 5(c) and (d). Therefore, on-chip multi-chromatic Raman lasing sources are demonstrated on high-Q square modes, by leveraging the novel broad-bandwidth phase matching approach, the high mode field overlap factors of the square modes.

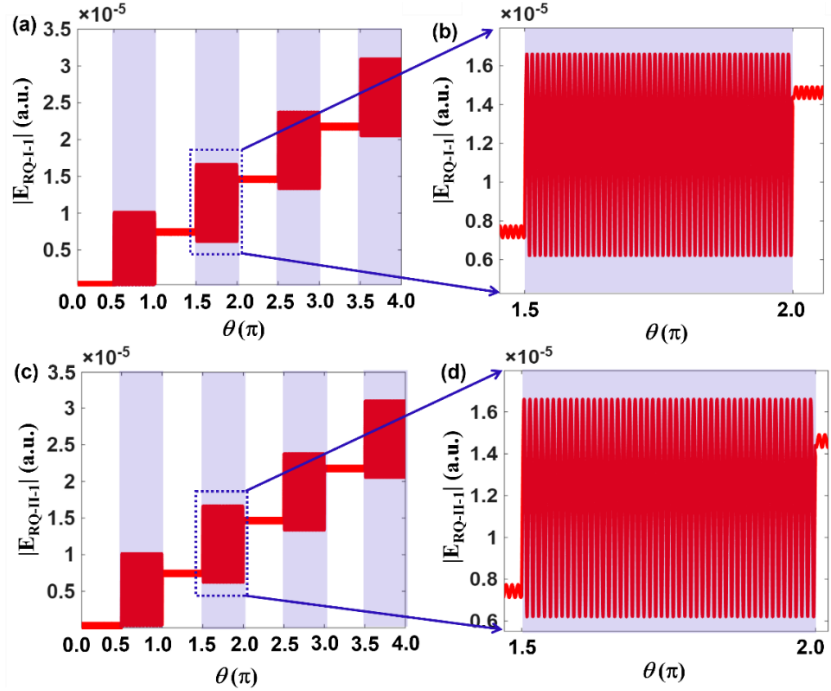

Fig. 5. (a) Stepped growth of the 797.4 nm Raman-quadratic signal amplitude versus azimuthal angle $\theta$ within 2 cavity-cycles. (b) Periodic oscillation and net gain of the 797.4 nm Raman-quadratic signal at the azimuthal angle section from $3\pi/2$ to $2\pi$. (c) Stepped growth of the 528.4 nm cascaded Raman-quadratic signal amplitude versus azimuthal angle $\theta$ (within 2 cavity-cycles. (d) Periodic oscillation and net gain of 528.4 nm signal at the azimuthal angle section from $3\pi/2$ to $2\pi$.

## 3. Conclusion

In summary, we demonstrated the generation of high-conversion-efficiency SRS based on polygon modes for the first time. Benefiting from the synergistic effect of three factors—high modal overlap of polygon modes, realization of natural quasi-phase matching, and an ultra-high intrinsic Q factor of up to $8.83\times10^6$, the conversion efficiency of this SRS reaches 65% in the forward direction, exceeding the current record of the highest conversion efficiency reported so far on the TFLN platform. Furthermore, through the combination of SRS with second-order nonlinear processes, we demonstrated multichromatic Raman-quadratic laser generation spanning from the wavelength range from the visible (528 nm) to the near-infrared (1816 nm). The absolute conversion efficiency for the 797 nm Raman-quadratic laser reached 1.33% at a pump power of only 1.07 mW, representing state-of-the-art performance.

Our results establish polygon modes as a versatile platform for cascaded nonlinear processes involving both second-order ($\chi^{(2)}$) and third-order ($\chi^{(3)}$) nonlinear optical processes, completing the nonlinear optical framework for this unique cavity mode configuration. The high modal overlap factor and natural quasi-phase matching condition provided by polygon modes are expected to facilitate higher conversion efficiencies in various other nonlinear phenomena. This work opens new avenues for the development of on-chip high-efficiency classical and quantum nonlinear light sources, with potential applications in quantum information processing, precision metrology, and optical communications.

**Acknowledgments.** We thank Peking university Yangtze delta institute of optoelectronics for providing the laser phase noise measurement system (iFN5000).